\documentclass[preprint,11pt]{elsarticle}

\usepackage{amssymb}

\usepackage{hyperref}
\usepackage{ulem}

\biboptions{sort&compress}

\usepackage{float}
\usepackage{fullpage}
\usepackage{color}

\restylefloat{table}

\journal{arXiv}

\begin{document}

\begin{frontmatter}



\title{Version 2.0.0 - SPARC: Simulation Package for Ab-initio Real-space Calculations \vspace{-2.5mm}}


\author[label1]{Boqin Zhang}
\author[label1]{Xin Jing}
\author[label1]{Qimen Xu}
\author[label1]{Shashikant Kumar}
\author[label3]{Abhiraj Sharma}
\author[label2]{Lucas Erlandson}
\author[label1]{Sushree Jagriti Sahoo}
\author[label2]{Edmond Chow}
\author[label1]{Andrew J. Medford}
\author[label3]{John E. Pask}
\author[label1,label2]{Phanish Suryanarayana\corref{cor1}}
\ead{phanish.suryanarayana@ce.gatech.edu}
\cortext[cor1]{corresponding author}
\address[label1]{College of Engineering, Georgia Institute of Technology, Atlanta, GA 30332, USA}
\address[label2]{College of Computing, Georgia Institute of Technology, Atlanta, GA 30332, USA}
\address[label3]{Physics Division, Lawrence Livermore National Laboratory, Livermore, CA 94550, USA}

\begin{abstract}
SPARC is an accurate, efficient, and scalable real-space electronic structure code for performing ab initio Kohn-Sham density functional theory calculations. Version 2.0.0 of the software provides increased efficiency, and includes spin-orbit coupling, dispersion interactions, and advanced semilocal as well as hybrid exchange-correlation functionals, where it outperforms state-of-the-art planewave codes by an order of magnitude and more, with increasing advantages as the number of processors is increased. These new features further expand the range of physical applications amenable to first principles investigation.
\end{abstract}

\begin{keyword}
Kohn–Sham Density Functional Theory \sep Electronic structure \sep Relativistic effects \sep Dispersion interactions \sep Meta-GGA exchange-correlation \sep Hybrid exchange-correlation



\end{keyword}
\end{frontmatter}

\section*{Metadata}
\label{metadata}
\vspace{-5.0mm}
\begin{table}[H]
\begin{tabular}{|l|p{6.5cm}|p{6.5cm}|}
\hline
C1 & Current code version & v2.0.0 \\
\hline
C2 & Permanent link to code/repository used for this code version & \url{https://github.com/SPARC-X/SPARC/releases/tag/v2.0.0} \\
\hline
C3 & Permanent link to Reproducible Capsule & N/A\\
\hline
C4 & Legal Code License   & GNU General Public License v3.0 \\
\hline
C5 & Code versioning system used & Git\\
\hline
C6 & Software code languages, tools, and services used & C, make, Python (for tests)\\
\hline
C7 & Compilation requirements, operating environments \& dependencies & Unix, Linux or MacOS; MPI, BLAS, LAPACK, ScaLAPACK (optional), MKL (optional), FFTW (optional)\\
\hline
C8 & If available Link to developer documentation/manual & \url{https://github.com/SPARC-X/SPARC/tree/master/doc} \\
\hline
C9 & Support email for questions & \url{phanish.s@gmail.com}\\
\hline
\end{tabular}
\end{table}

%





\section{Introduction} \label{Sec:Introduction}

Over the past two decades, Kohn-Sham density functional theory (DFT) \cite{kohn1965self, hohenberg1964inhomogeneous} has established itself as one of the cornerstones of materials and chemical sciences research. However, Kohn-Sham  calculations are associated with significant computational cost, scaling cubically with system size, severely limiting the length and time scales accessible to such a rigorous first principles investigation. The planewave pseudopotential method \cite{martin2020electronic} has been among the most widely used techniques for the solution of the Kohn-Sham problem \cite{VASP, CASTEP, ABINIT, Espresso, CPMD, DFT++,gygi2008architecture, valiev2010nwchem}. However, it is limited to periodic boundary conditions due to the underlying Fourier basis, whose global nature also hampers scalability on parallel computing platforms. This has motivated the development of alternative approaches  employing systematically improvable, localized representations \cite{becke1989basis, chelikowsky1994finite, genovese2008daubechies, seitsonen1995real, white1989finite, iwata2010massively, tsuchida1995electronic, xu2018discrete, Phanish2011, Phanish2010, ONETEP, CONQUEST, MOTAMARRI2020106853, OCTOPUS, briggs1996real, fattebert1999finite, shimojo2001linear, ghosh2017sparcI, ghosh2017sparc2, arias1999wav, pask2005femeth, lin2012adaptive}, with finite-difference methods perhaps the most mature and widely used to date.

SPARC \cite{xu2021sparc} is a real-space electronic structure code for performing Kohn-Sham DFT calculations, where all quantities of interest, such as densities, potentials, and wavefunctions, are discretized on a uniform grid using the finite-difference approximation, enabling systematic control of convergence with a single parameter. It can accommodate both Dirichlet and Bloch-periodic boundary conditions, enabling the accurate and efficient treatment of finite, semi-infinite, and bulk 3D systems. SPARC has been extensively validated and benchmarked against established planewave codes. In particular, SPARC is able to efficiently leverage moderate and large-scale computational resources alike, efficiently scaling to thousands of processors in regular operation, and demonstrating order-of-magnitude speedups in time to solution relative to state-of-the-art planewave codes, with increasing advantages as the number of processors is increased.


\section{Description of the software update}
\label{Description of the software-update}
 
SPARC is designed to be portable and straightforward to install, use, and modify. In particular, external dependencies are limited to industry standard BLAS, LAPACK/ScaLAPACK, and MPI.  It can perform spin-unpolarized as well as spin-polarized calculations, i.e., with fixed cell and ionic positions; geometry optimizations with respect to atom positions and/or cell; and ab initio molecular dynamics (AIMD) simulations, all using norm-conserving pseudopotentials \cite{hamann2013optimized,troullier1991efficient} with a range of local exchange-correlation functionals. Version 2.0.0 further expands the range of physical applications amenable to first principles investigation by further increasing efficiency and including spin-orbit coupling, dispersion interactions, and advanced semilocal as well as hybrid exchange–correlation functionals beyond the generalized gradient approximation (GGA) \cite{martin2020electronic}, as described below.

\begin{itemize}
    \item \emph{Spin-orbit coupling (SOC)}: SOC refers to the relativistic electromagnetic interaction between the  electron's spin and orbital angular momenta \cite{martin2020electronic}. SOC becomes more important as the atomic number increases, having a significant impact on the spectra of systems with heavy atoms.  SOC has been implemented in SPARC through  relativistic norm-conserving pseudopotentials \cite{kleinman1980relativistic}, using the real-space formalism presented in Ref.~\cite{naveh2007real}. 
\item \emph{Dispersion interactions}: The van der Waals (vdW) dispersion interaction, which is a correlation effect, represents the coupling between different parts of the system due to electronic charge fluctuations \cite{martin2020electronic}. This long-range interaction becomes increasingly important in molecular and layered systems. Dispersion interactions have been implemented in SPARC through the pairwise DFT-D3 correction \cite{grimme2010consistent} and the nonlocal vdW density functional (vdW-DF) \cite{dion2004van, lee2010higher}, for which we adopt the method presented in Refs.~\cite{roman2009efficient, thonhauser2015spin}.
\item \emph{Meta-GGA exchange-correlation functionals}: Residing on the third rung of Jacob's ladder \cite{perdew2001jacob}, one rung above GGA, meta-GGA exchange-correlation functionals include a semilocal term that is dependent on the kinetic energy density, in addition to terms that are dependent on the electron density and its gradient which form the basis of GGA  \cite{martin2020electronic}. meta-GGA  is implemented in SPARC through the recently developed SCAN functional, which satisfies  all seventeen of the constraints  known on the universal exchange-correlation functional \cite{sun2015strongly}.
\item \emph{Hybrid exchange-correlation functionals}: Residing on the fourth rung of Jacob's ladder, one rung above meta-GGA, hybrid functionals include a fraction of the nonlocal Hartree-Fock exact exchange energy, in addition to the terms found in meta-GGA and GGA. Hybrid functionals have been implemented in SPARC through the PBE0 \cite{adamo1999toward} and HSE \cite{heyd2003hybrid} functionals,  using the methods presented in Refs.~\cite{lin2016adaptively,spencer2008efficient,gygi1986self} for evaluating the exact exchange contribution.
\end{itemize}    
In addition to the above functionalities, nonlinear core corrections (NLCC) --- in which nonlinearity in the exchange-correlation functional is taken into account within the pseudopotential formalism --- have been implemented; an extensive automated testing framework that includes examples with a wide range of system compositions, configurations, and dimensionalities has been developed;  the SPMS table of pseudopotentials \cite{shojaei2023soft} --- accurate, transferable, and soft optimized norm-conserving Vanderbilt (ONCV) pseudopotentials \cite{hamann2013optimized} with NLCC —-- has been incorporated into the distribution; and  the overall implementation has been refined: optimized application of boundary conditions during stencil operations, optimized input parameters for the dense eigensolvers in ScaLAPACK,  improved defaults for parallelization in the presence of Brillouin zone integration, and reformulation of the energy as well as forces to enable faster convergence within the self-consistent iteration, to further increase efficiency and parallel scaling.

\begin{figure}[htbp!]
\centering
\includegraphics[width=0.98\textwidth]{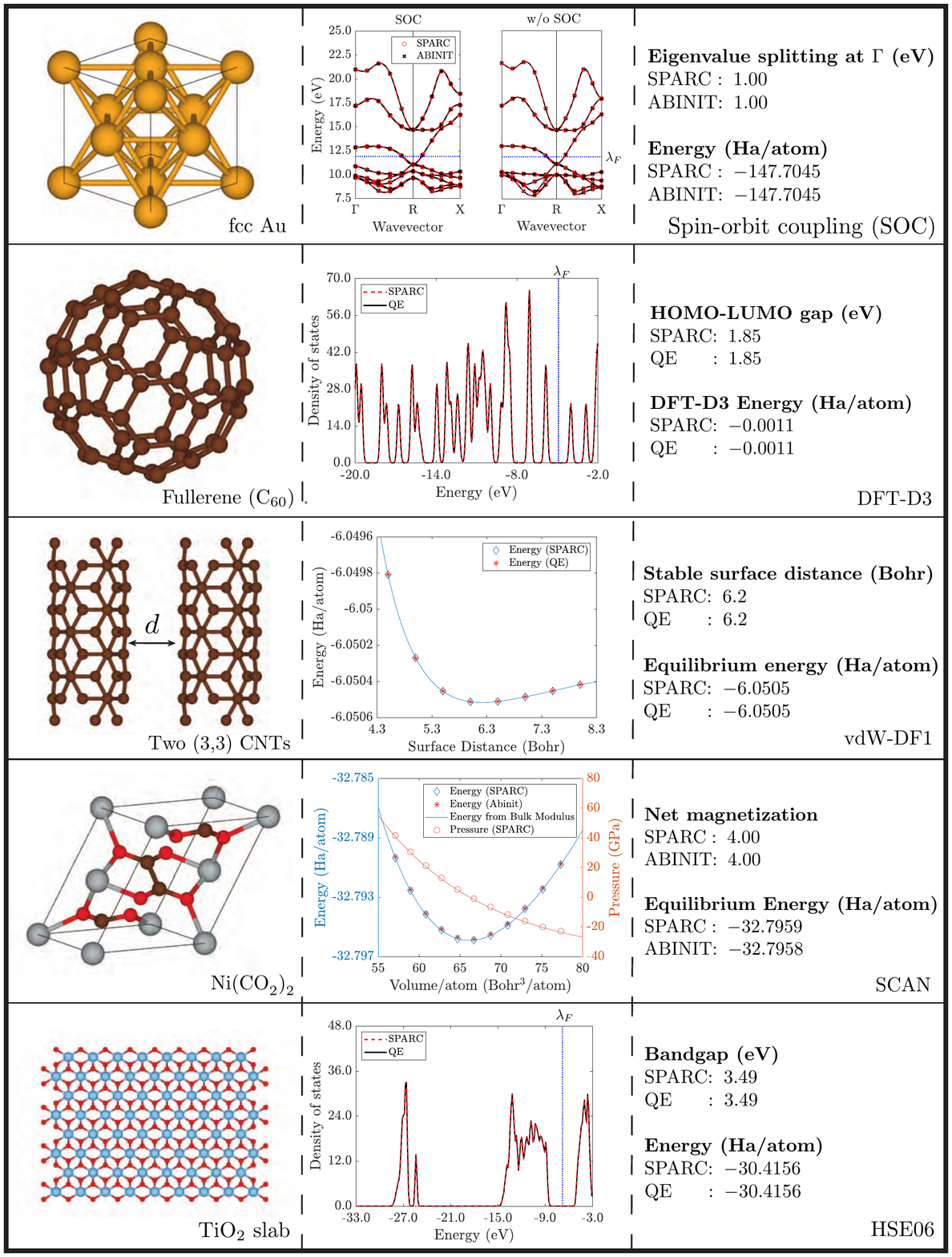}
\caption{Examples demonstrating the accuracy of the new features in SPARC v2.0.0.}
\label{fig1}
\end{figure}

In Fig.~\ref{fig1}, we demonstrate the accuracy of the major new functionalities in SPARC through representative examples. In particular, we compare against the established planewave codes ABINIT v7.6.2 \cite{Gonze2020} and Quantum Espresso (QE) v7.1 \cite{Espresso}. Unless otherwise specified, we perform static calculations using ONCV pseudopotentials from the SPMS set and a 12th-order finite difference approximation for discretization. To demonstrate the accuracy of SPARC, we consider the following systems: (i) 4-atom cell of face-centered cubic (fcc) gold with PBE exchange-correlation, SOC using the relativistic ONCV pseudopotential from the PseudoDOJO set \cite{van2018pseudodojo}, $10\times10\times10$ grid for Brillouin zone integration, and real-space grid spacing of 0.15 bohr; (ii) fullerene molecule with PBE exchange-correlation, dispersion interactions through DFT-D3, and real-space grid spacing of 0.24 bohr; (iii) two 12-atom (3,3) carbon nanotubes with vdW-DF exchange-correlation, $10$ grid points for Brillouin zone integration, and real-space grid spacing of 0.15 bohr; (iv) 14-atom cell of bulk Ni(CO$_2$)$_2$ with SCAN meta-GGA exchange-correlation, ONCV pseudopotentials without NLCC distributed with the ONCVPSP pseudopotential generation code \cite{oncvpsp} (SPARC v2.0.0 and the chosen versions of ABINIT as well as QE do not support SCAN with NLCC), $2\times2\times2$ grid for Brillouin zone integration, and  real-space grid spacing of 0.15 bohr; and (v) 24-atom cell of TiO$_2$ with HSE hybrid exchange-correlation, $6\times6$ grid for Brillouin zone integration, and real-space grid spacing of 0.25 bohr. As shown in Fig.~\ref{fig1}, SPARC shows excellent agreement with ABINIT as well as QE, where agreement can be increased as desired by refining the real-space grid.

\begin{figure}[htbp!]
\centering
\includegraphics[width=0.97\textwidth]{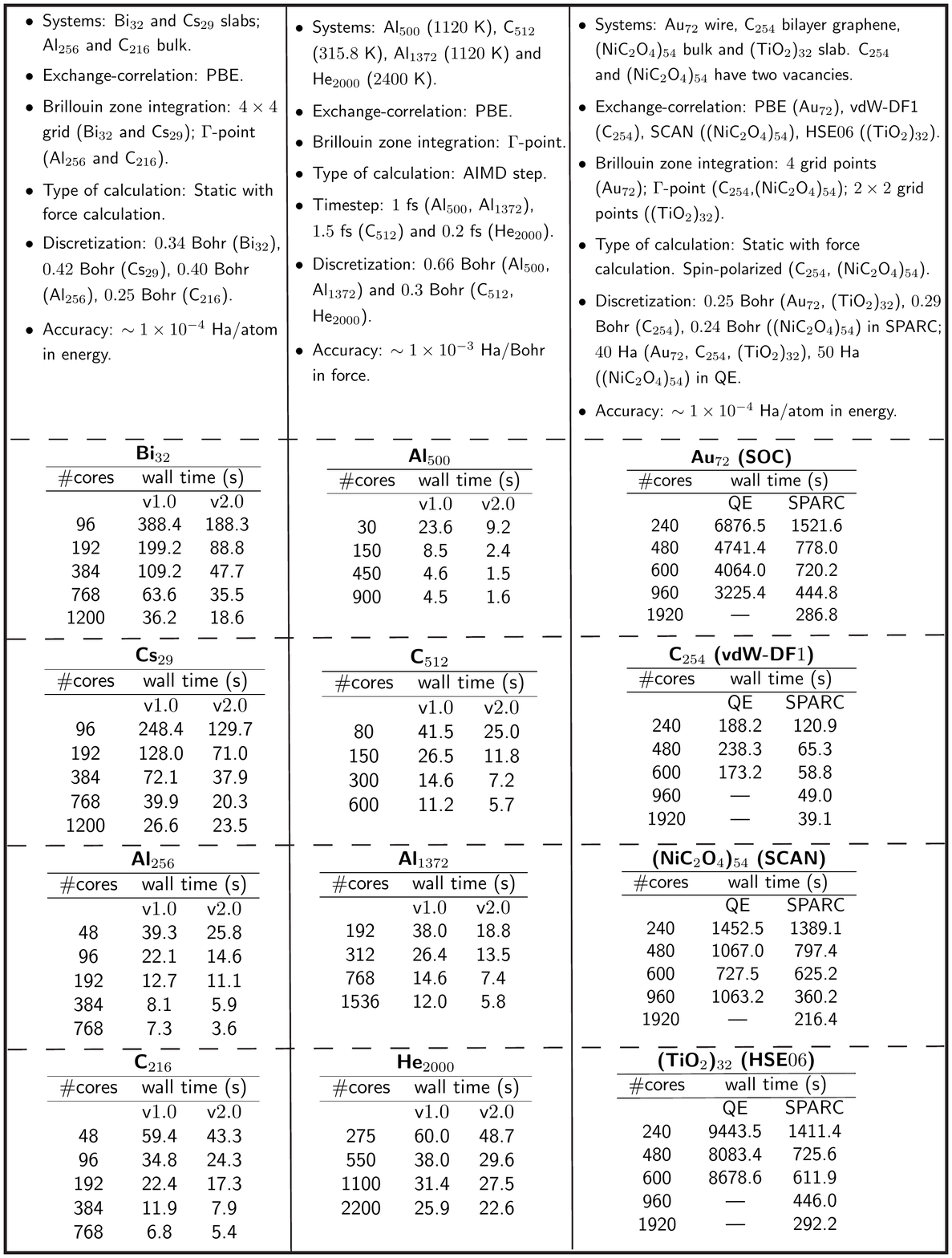}
\caption{Examples demonstrating the performance of SPARC v2.0.0.}
\label{fig2}
\end{figure}

Next, we demonstrate the efficiency and scaling of SPARC v2.0.0 through representative examples, as shown in Fig.~\ref{fig2}.  For new features, we compare against the state-of-the-art parallel planewave code  QE v7.1, whereas for previously implemented features, we compare against SPARC v1.0.0, which has already been benchmarked  against QE in Ref.~\cite{xu2021sparc}. Unless otherwise specified, we use ONCV pseudopotentials from the SPMS set and a 12th-order finite difference approximation for discretization. We use a relativistic ONCV pseudopotential from the PseudoDOJO set for the SOC calculation, and ONCV pseudopotentials without NLCC distributed with the ONCVPSP pseudopotential generation code for the SCAN calculation. In both SPARC and QE, we use the default parallelization settings. All calculations are performed on the Phoenix cluster at Georgia Tech. comprised  of Dual Intel Xeon Gold 6226 CPUs @ 2.7 GHz (24 cores/node), DDR4-2933 MHz DRAM, and Infiniband 100HDR interconnect. Additional details regarding the systems selected, which include a wide range of sizes, the type of calculation performed, and computational parameters chosen for this study are shown in Fig.~\ref{fig2}, the raw data for which is provided in Ref.~\cite{zhang2023SI}. It can be seen that SPARC demonstrates up to an order-of-magnitude speedup in time to solution with respect to QE for the new features, with increasing advantages as the number of processors is increased. Notably, the largest speedup occurs for the hybrid functional, due mainly to the superior scalability of domain decomposition in real-space vs. planewave implementations. Furthermore, it can be seen that SPARC v2.0.0 is up to a factor of two faster than v1.0.0, which is itself an order of magnitude faster than QE \cite{xu2021sparc}.

\section{Impact and future development}

Kohn–Sham DFT calculations occupy a substantial fraction of the world's high-performance computing resources every day \cite{nersc2014, lanlcomp2015}. Since most of these calculations are performed using established planewave DFT codes \cite{VASP, CASTEP, Espresso, ABINIT, CPMD, valiev2010nwchem}, any new code that consistently outperforms these state-of-the-art codes stands to have  immediate and significant impact. The new functionalities in SPARC v2.0.0 allow for higher fidelity first principles simulations  than possible with the previous version. Given that some of these features are computationally intensive, e.g., hybrid exchange-correlation functionals, the superior efficiency and scalability of SPARC relative to  established planewave alternatives, demonstrating order of magnitude and more speedups, with increasing advantages as the number of processors is increased, stands to enable a number of physical applications  that were previously beyond reach, as  evidenced by  recent publications employing SPARC \cite{sahoo2022ab, sahoo2024self, zeng2023phase, pathrudkar2023electronic}.

Future releases of SPARC will include the following features, which while being under development in SPARC, have resulted in a number of publications \cite{sharma2023gpu, suryanarayana2018sqdft, gavini2022roadmap, bethkenhagen2023properties, zhang2019equation, wu2021development, sharma2021real, ghosh2019symmetry,  kumar2020bending, kumar2022bending, codony2021transversal, kumar2021flexoelectricity, bhardwaj2022strain, bhardwaj2021torsionala, bhardwaj2021torsionalb, bhardwaj2023ab, bhardwaj2022elastic, ghosh2016higher, suryanarayana2014augmented, kumar2023kohn, kumar2024fly}: the Discrete Discontinuous Basis Projection (DDBP) method \cite{xu2018discrete}, which brings down time to solution for large problems; GPU acceleration to bring down time to solution further still \cite{sharma2023gpu}; the linear scaling Spectral Quadrature (SQ) method \cite{suryanarayana2018sqdft, pratapa2016spectral, suryanarayana2013spectral}, which has enabled system sizes as large as a million atoms \cite{gavini2022roadmap} and has found a number of applications in warm dense matter \cite{bethkenhagen2023properties, zhang2019equation, wu2021development}; cyclic+helical symmetry-adapted DFT \cite{sharma2021real, ghosh2019symmetry, banerjee2016cyclic}, referred to as Cyclix-DFT \cite{sharma2021real}, which has found a number of applications in the study of 1D and 2D nanomaterials subject to mechanical deformations \cite{kumar2020bending, kumar2022bending, codony2021transversal, kumar2021flexoelectricity, bhardwaj2022strain, bhardwaj2021torsionala, bhardwaj2021torsionalb, bhardwaj2023ab, bhardwaj2022elastic};  real-space Density Functional Perturbation Theory (DFPT), which will enable the efficient calculation of the response to perturbations, e.g., phonons \cite{sharma2023calculation}; correlation energy within the Random Phase Approximation \cite{martin2020electronic}; orbital-free DFT \cite{ghosh2016higher, suryanarayana2014augmented}, which has applications in the development of machine learning techniques \cite{thapa2023assessing, kumar2022accurate}; and on-the-fly machine-learned force-fields \cite{kumar2023kohn, kumar2024fly}, with the capability to treat many-element systems using the Gaussian multipole featurization scheme \cite{lei2022universal}. The development of such features is being accelerated by prototyping in the \texttt{MATLAB} version of SPARC, referred to as M-SPARC \cite{zhang2023version, xu2020m, huang2023formation}.

\section*{Acknowledgements}
{J.E.P, P.S., B.Z., and X.J. gratefully acknowledge support from U.S. Department of Energy (DOE), National Nuclear Security Administration (NNSA): Advanced Simulation and Computing (ASC) Program at LLNL. The authors also gratefully acknowledge support from grant DE-SC0019410 funded by the U.S. Department of Energy, Office of Science. This work was performed in part under the auspices of the U.S. Department of Energy by Lawrence Livermore National Laboratory under Contract DE-AC52-07NA27344.



\bibliographystyle{elsarticle-num} 

\begin{thebibliography}{10}
\expandafter\ifx\csname url\endcsname\relax
  \def\url#1{\texttt{#1}}\fi
\expandafter\ifx\csname urlprefix\endcsname\relax\def\urlprefix{URL }\fi
\expandafter\ifx\csname href\endcsname\relax
  \def\href#1#2{#2} \def\path#1{#1}\fi

\bibitem{kohn1965self}
W.~Kohn, L.~J. Sham, Self-consistent equations including exchange and
  correlation effects, Physical Review 140~(4A) (1965) A1133.

\bibitem{hohenberg1964inhomogeneous}
P.~Hohenberg, W.~Kohn, Inhomogeneous electron gas, Physical Review 136~(3B)
  (1964) 864.

\bibitem{martin2020electronic}
R.~M. Martin, Electronic structure: basic theory and practical methods,
  Cambridge University Press, 2020.

\bibitem{VASP}
G.~Kresse, J.~Furthm\"uller, Efficient iterative schemes for ab initio
  total-energy calculations using a plane-wave basis set, Physical Review B
  54~(16) (1996) 11169--11186.

\bibitem{CASTEP}
S.~J. Clark, M.~D. Segall, C.~J. Pickard, P.~J. Hasnip, M.~I. Probert,
  K.~Refson, M.~C. Payne, {First principles methods using CASTEP}, Zeitschrift
  f{\"u}r Kristallographie-Crystalline Materials 220~(5/6) (2005) 567--570.

\bibitem{ABINIT}
X.~Gonze, J.~M. Beuken, R.~Caracas, F.~Detraux, M.~Fuchs, G.~M. Rignanese,
  L.~Sindic, M.~Verstraete, G.~Zerah, F.~Jollet, M.~Torrent, A.~Roy, M.~Mikami,
  P.~Ghosez, J.~Y. Raty, D.~C. Allan, First-principles computation of material
  properties: the {ABINIT} software project, Computational Materials Science 25
  (2002) 478--492(15).

\bibitem{Espresso}
P.~Giannozzi, S.~Baroni, N.~Bonini, M.~Calandra, R.~Car, C.~Cavazzoni,
  D.~Ceresoli, G.~L. Chiarotti, M.~Cococcioni, I.~Dabo, A.~D. Corso,
  S.~de~Gironcoli, S.~Fabris, G.~Fratesi, R.~Gebauer, U.~Gerstmann,
  C.~Gougoussis, A.~Kokalj, M.~Lazzeri, L.~Martin-Samos, N.~Marzari, F.~Mauri,
  R.~Mazzarello, S.~Paolini, A.~Pasquarello, L.~Paulatto, C.~Sbraccia,
  S.~Scandolo, G.~Sclauzero, A.~P. Seitsonen, A.~Smogunov, P.~Umari, R.~M.
  Wentzcovitch, {QUANTUM ESPRESSO: a modular and open-source software project
  for quantum simulations of materials}, Journal of Physics: Condensed Matter
  21~(39) (2009) 395502.

\bibitem{CPMD}
D.~Marx, J.~Hutter, Ab initio molecular dynamics: Theory and implementation,
  Modern methods and algorithms of quantum chemistry 1 (2000) 301--449.

\bibitem{DFT++}
S.~Ismail-Beigi, T.~A. Arias, New algebraic formulation of density functional
  calculation, Computer Physics Communications 128~(1-2) (2000) 1 -- 45.

\bibitem{gygi2008architecture}
F.~Gygi, {Architecture of Qbox: A scalable first-principles molecular dynamics
  code}, IBM Journal of Research and Development 52~(1.2) (2008) 137--144.

\bibitem{valiev2010nwchem}
M.~Valiev, E.~Bylaska, N.~Govind, K.~Kowalski, T.~Straatsma, H.~V. Dam,
  D.~Wang, J.~Nieplocha, E.~Apra, T.~Windus, W.~de~Jong, {NWChem: A
  comprehensive and scalable open-source solution for large scale molecular
  simulations}, Computer Physics Communications 181~(9) (2010) 1477 -- 1489.

\bibitem{becke1989basis}
A.~D. Becke, Basis-set-free density-functional quantum chemistry, International
  Journal of Quantum Chemistry 36~(S23) (1989) 599--609.

\bibitem{chelikowsky1994finite}
J.~R. Chelikowsky, N.~Troullier, Y.~Saad, Finite-difference-pseudopotential
  method: Electronic structure calculations without a basis, Physical Review
  Letters 72~(8) (1994) 1240--1243.
\newblock \href {http://dx.doi.org/10.1103/PhysRevLett.72.1240}
  {\path{doi:10.1103/PhysRevLett.72.1240}}.

\bibitem{genovese2008daubechies}
L.~Genovese, A.~Neelov, S.~Goedecker, T.~Deutsch, S.~A. Ghasemi, A.~Willand,
  D.~Caliste, O.~Zilberberg, M.~Rayson, A.~Bergman, R.~Schneider, Daubechies
  wavelets as a basis set for density functional pseudopotential calculations,
  The Journal of Chemical Physics 129~(1) (2008) 014109.

\bibitem{seitsonen1995real}
A.~P. Seitsonen, M.~J. Puska, R.~M. Nieminen, Real-space electronic-structure
  calculations: Combination of the finite-difference and conjugate-gradient
  methods, Physical Review B 51~(20) (1995) 14057.

\bibitem{white1989finite}
S.~R. White, J.~W. Wilkins, M.~P. Teter, Finite-element method for electronic
  structure, Physical Review B 39~(9) (1989) 5819.

\bibitem{iwata2010massively}
J.-I. Iwata, D.~Takahashi, A.~Oshiyama, T.~Boku, K.~Shiraishi, S.~Okada,
  K.~Yabana, A massively-parallel electronic-structure calculations based on
  real-space density functional theory, Journal of Computational Physics
  229~(6) (2010) 2339--2363.

\bibitem{tsuchida1995electronic}
E.~Tsuchida, M.~Tsukada, Electronic-structure calculations based on the
  finite-element method, Physical Review B 52~(8) (1995) 5573.

\bibitem{xu2018discrete}
Q.~Xu, P.~Suryanarayana, J.~E. Pask, Discrete discontinuous basis projection
  method for large-scale electronic structure calculations, The Journal of
  Chemical Physics 149~(9) (2018) 094104.

\bibitem{Phanish2011}
P.~Suryanarayana, K.~Bhattacharya, M.~Ortiz, A mesh-free convex approximation
  scheme for {Kohn}-{Sham} density functional theory, Journal of Computational
  Physics 230~(13) (2011) 5226 -- 5238.

\bibitem{Phanish2010}
P.~Suryanarayana, V.~Gavini, T.~Blesgen, K.~Bhattacharya, M.~Ortiz,
  Non-periodic finite-element formulation of {Kohn}-{Sham} density functional
  theory, Journal of the Mechanics and Physics of Solids 58~(2) (2010) 256 --
  280.

\bibitem{ONETEP}
C.-K. Skylaris, P.~D. Haynes, A.~A. Mostofi, M.~C. Payne, Introducing {ONETEP}:
  Linear-scaling density functional simulations on parallel computers, The
  Journal of Chemical Physics 122~(8) (2005) 084119.

\bibitem{CONQUEST}
D.~R. Bowler, R.~Choudhury, M.~J. Gillan, T.~Miyazaki, Recent progress with
  large-scale ab initio calculations: the {CONQUEST} code, physica status
  solidi (b) 243~(5) (2006) 989--1000.

\bibitem{MOTAMARRI2020106853}
P.~Motamarri, S.~Das, S.~Rudraraju, K.~Ghosh, D.~Davydov, V.~Gavini, {DFT-FE
  --- A massively parallel adaptive finite-element code for large-scale density
  functional theory calculations}, Computer Physics Communications 246 (2020)
  106853.

\bibitem{OCTOPUS}
A.~Castro, H.~Appel, M.~Oliveira, C.~A. Rozzi, X.~Andrade, F.~Lorenzen,
  M.~A.~L. Marques, E.~K.~U. Gross, A.~Rubio, {octopus: a tool for the
  application of time-dependent density functional theory}, {Physica Status
  Solidi B-Basic Solid State Physics} {243}~({11}) ({2006}) {2465--2488}.

\bibitem{briggs1996real}
E.~L. Briggs, D.~J. Sullivan, J.~Bernholc, Real-space multigrid-based approach
  to large-scale electronic structure calculations, Physical Review B 54 (1996)
  14362--14375.

\bibitem{fattebert1999finite}
J.-L. Fattebert, Finite difference schemes and block {Rayleigh} quotient
  iteration for electronic structure calculations on composite grids, Journal
  of Computational Physics 149~(1) (1999) 75 -- 94.

\bibitem{shimojo2001linear}
F.~Shimojo, R.~K. Kalia, A.~Nakano, P.~Vashishta, Linear-scaling
  density-functional-theory calculations of electronic structure based on
  real-space grids: design, analysis, and scalability test of parallel
  algorithms, Computer Physics Communications 140~(3) (2001) 303 -- 314.

\bibitem{ghosh2017sparcI}
S.~Ghosh, P.~Suryanarayana, {SPARC: Accurate and efficient finite-difference
  formulation and parallel implementation of density functional theory:
  Isolated clusters}, Computer Physics Communications 212 (2017) 189--204.

\bibitem{ghosh2017sparc2}
S.~Ghosh, P.~Suryanarayana, {SPARC: Accurate and efficient finite-difference
  formulation and parallel implementation of density functional theory:
  Extended systems}, Computer Physics Communications 216 (2017) 109--125.

\bibitem{arias1999wav}
T.~A. Arias, Multiresolution analysis of electronic structure: semicardinal and
  wavelet bases, Reviews of Modern Physics 71~(1) (1999) 267--311.

\bibitem{pask2005femeth}
J.~E. Pask, P.~A. Sterne, Finite element methods in ab initio electronic
  structure calculations, Modelling and Simulation in Materials Science and
  Engineering 13 (2005) R71--R96.

\bibitem{lin2012adaptive}
L.~Lin, J.~Lu, L.~Ying, E.~Weinan, Adaptive local basis set for {Kohn}--{Sham}
  density functional theory in a discontinuous {Galerkin} framework i: Total
  energy calculation, Journal of Computational Physics 231~(4) (2012)
  2140--2154.

\bibitem{xu2021sparc}
Q.~Xu, A.~Sharma, B.~Comer, H.~Huang, E.~Chow, A.~J. Medford, J.~E. Pask,
  P.~Suryanarayana, {SPARC}: Simulation package for ab-initio real-space
  calculations, SoftwareX 15 (2021) 100709.

\bibitem{hamann2013optimized}
D.~Hamann, {Optimized norm-conserving Vanderbilt pseudopotentials}, Physical
  Review B 88~(8) (2013) 085117.

\bibitem{troullier1991efficient}
N.~Troullier, J.~L. Martins, Efficient pseudopotentials for plane-wave
  calculations, Physical Review B 43~(3) (1991) 1993.

\bibitem{kleinman1980relativistic}
L.~Kleinman, Relativistic norm-conserving pseudopotential, Physical Review B
  21~(6) (1980) 2630.

\bibitem{naveh2007real}
D.~Naveh, L.~Kronik, M.~L. Tiago, J.~R. Chelikowsky, Real-space pseudopotential
  method for spin-orbit coupling within density functional theory, Physical
  Review B 76~(15) (2007) 153407.

\bibitem{grimme2010consistent}
S.~Grimme, J.~Antony, S.~Ehrlich, H.~Krieg, {A consistent and accurate ab
  initio parametrization of density functional dispersion correction (DFT-D)
  for the 94 elements H-Pu}, The Journal of Chemical Physics 132~(15) (2010)
  154104.

\bibitem{dion2004van}
M.~Dion, H.~Rydberg, E.~Schr{\"o}der, D.~C. Langreth, B.~I. Lundqvist, {Van der
  Waals density functional for general geometries}, Physical Review Letters
  92~(24) (2004) 246401.

\bibitem{lee2010higher}
K.~Lee, {\'E}.~D. Murray, L.~Kong, B.~I. Lundqvist, D.~C. Langreth,
  {Higher-accuracy van der Waals density functional}, Physical Review B 82~(8)
  (2010) 081101.

\bibitem{roman2009efficient}
G.~Rom{\'a}n-P{\'e}rez, J.~M. Soler, {Efficient implementation of a van der
  Waals density functional: application to double-wall carbon nanotubes},
  Physical Review Letters 103~(9) (2009) 096102.

\bibitem{thonhauser2015spin}
T.~Thonhauser, S.~Zuluaga, C.~Arter, K.~Berland, E.~Schr{\"o}der, P.~Hyldgaard,
  Spin signature of nonlocal correlation binding in metal-organic frameworks,
  Physical Review Letters 115~(13) (2015) 136402.

\bibitem{perdew2001jacob}
J.~P. Perdew, K.~Schmidt, Jacob’s ladder of density functional approximations
  for the exchange-correlation energy, AIP Conference Proceedings 577 (2001)
  1--20.

\bibitem{sun2015strongly}
J.~Sun, A.~Ruzsinszky, J.~P. Perdew, Strongly constrained and appropriately
  normed semilocal density functional, Physical Review Letters 115~(3) (2015)
  036402.

\bibitem{adamo1999toward}
C.~Adamo, V.~Barone, {Toward reliable density functional methods without
  adjustable parameters: The PBE0 model}, The Journal of Chemical Physics
  110~(13) (1999) 6158--6170.

\bibitem{heyd2003hybrid}
J.~Heyd, G.~E. Scuseria, M.~Ernzerhof, {Hybrid functionals based on a screened
  Coulomb potential}, The Journal of Chemical Physics 118~(18) (2003)
  8207--8215.

\bibitem{lin2016adaptively}
L.~Lin, Adaptively compressed exchange operator, Journal of Chemical Theory and
  Computation 12~(5) (2016) 2242--2249.

\bibitem{spencer2008efficient}
J.~Spencer, A.~Alavi, {Efficient calculation of the exact exchange energy in
  periodic systems using a truncated Coulomb potential}, Physical Review B
  77~(19) (2008) 193110.

\bibitem{gygi1986self}
F.~Gygi, A.~Baldereschi, {Self-consistent Hartree-Fock and screened-exchange
  calculations in solids: Application to silicon}, Physical Review B 34~(6)
  (1986) 4405.

\bibitem{shojaei2023soft}
M.~F. Shojaei, J.~E. Pask, A.~J. Medford, P.~Suryanarayana, Soft and
  transferable pseudopotentials from multi-objective optimization, Computer
  Physics Communications 283 (2023) 108594.

\bibitem{Gonze2020}
X.~Gonze, B.~Amadon, G.~Antonius, F.~Arnardi, L.~Baguet, J.-M. Beuken,
  J.~Bieder, F.~Bottin, J.~Bouchet, E.~Bousquet, N.~Brouwer, F.~Bruneval,
  G.~Brunin, T.~Cavignac, J.-B. Charraud, W.~Chen, M.~Côté, S.~Cottenier,
  J.~Denier, G.~Geneste, P.~Ghosez, M.~Giantomassi, Y.~Gillet, O.~Gingras,
  D.~R. Hamann, G.~Hautier, X.~He, N.~Helbig, N.~Holzwarth, Y.~Jia, F.~Jollet,
  W.~Lafargue-Dit-Hauret, K.~Lejaeghere, M.~A. Marques, A.~Martin, C.~Martins,
  H.~P. Miranda, F.~Naccarato, K.~Persson, G.~Petretto, V.~Planes, Y.~Pouillon,
  S.~Prokhorenko, F.~Ricci, G.-M. Rignanese, A.~H. Romero, M.~M. Schmitt,
  M.~Torrent, M.~J. {van Setten}, B.~{Van Troeye}, M.~J. Verstraete, G.~Zérah,
  J.~W. Zwanziger, The {ABINIT} project: Impact, environment and recent
  developments, Computer Physics Communications 248 (2020) 107042.

\bibitem{van2018pseudodojo}
M.~J. van Setten, M.~Giantomassi, E.~Bousquet, M.~J. Verstraete, D.~R. Hamann,
  X.~Gonze, G.-M. Rignanese, {The PseudoDojo: Training and grading a 85 element
  optimized norm-conserving pseudopotential table}, Computer Physics
  Communications 226 (2018) 39--54.

\bibitem{oncvpsp}
D.~R. Hamann, {ONCVPSP} pseudopotential generation code,
  \url{www.mat-simresearch.com}, accessed: 2023-05-08.

\bibitem{zhang2023SI}
B.~Zhang, X.~Jing, Q.~Xu, S.~Kumar, A.~Sharma, L.~Erlandson, S.~J. Sahoo,
  E.~Chow, A.~J. Medford, J.~E. Pask, P.~Suryanarayana, {Supporting Information
  for Version 2.0.0 - SPARC: Simulation Package for Ab-initio Real-space
  Calculations}, Mendeley Data V1.
\newblock \href {http://dx.doi.org/10.17632/mvsc6cznrm.1}
  {\path{doi:10.17632/mvsc6cznrm.1}}.

\bibitem{nersc2014}
B.~Austin, W.~Bhimji, T.~Butler, J.~Deslippe, {2014 NERSC workload analysis},
  \url{http://portal.nersc.gov/project/mpccc/baustin/NERSC_2014_Workload_Analysis_v1.1.pdf}.

\bibitem{lanlcomp2015}
L.~J. Vernon, {IC Application Performance Team analysis, as part of IC Knights
  Special Project}, {Tech. Rep. LANL, 2015}.

\bibitem{sahoo2022ab}
S.~J. Sahoo, X.~Jing, P.~Suryanarayana, A.~J. Medford, Ab-initio investigation
  of finite size effects in rutile titania nanoparticles with semilocal and
  nonlocal density functionals, The Journal of Physical Chemistry C 126~(4)
  (2022) 2121--2130.

\bibitem{sahoo2024self}
S.~J. Sahoo, Q.~Xu, X.~Lei, D.~Staros, G.~R. Iyer, B.~Rubenstein,
  P.~Suryanarayana, A.~Medford, Self-consistent convolutional density
  functional approximations: Application to adsorption at metal surfaces,
  ChemPhysChem (2024) e202300688.

\bibitem{zeng2023phase}
C.~Zeng, S.~J. Sahoo, A.~J. Medford, A.~A. Peterson, Phase stability of
  large-size nanoparticle alloy catalysts at ab initio quality using a
  nearsighted force-training approach, The Journal of Physical Chemistry C
  127~(50) (2023) 24360--24372.

\bibitem{pathrudkar2023electronic}
S.~Pathrudkar, P.~Thiagarajan, S.~Agarwal, A.~S. Banerjee, S.~Ghosh, Electronic
  structure prediction of multi-million atom systems through uncertainty
  quantification enabled transfer learning, arXiv preprint arXiv:2308.13096.

\bibitem{sharma2023gpu}
A.~Sharma, A.~Metere, P.~Suryanarayana, L.~Erlandson, E.~Chow, J.~E. Pask, {GPU
  acceleration of local and semilocal density functional calculations in the
  SPARC electronic structure code}, The Journal of Chemical Physics 158~(20)
  (2023) 204117.

\bibitem{suryanarayana2018sqdft}
P.~Suryanarayana, P.~P. Pratapa, A.~Sharma, J.~E. Pask, {SQDFT: Spectral
  Quadrature method for large-scale parallel O(N) Kohn--Sham calculations at
  high temperature}, Computer Physics Communications 224 (2018) 288--298.

\bibitem{gavini2022roadmap}
V.~Gavini, S.~Baroni, V.~Blum, D.~R. Bowler, A.~Buccheri, J.~R. Chelikowsky,
  S.~Das, W.~Dawson, P.~Delugas, M.~Dogan, C.~Draxl, G.~Galli, L.~Genovese,
  P.~Giannozzi, M.~Giantomassi, X.~Gonze, M.~Govoni, F.~Gygi, A.~Gulans, J.~M.
  Herbert, S.~Kokott, T.~D. K{\"u}hne, K.-H. Liou, T.~Miyazaki, P.~Motamarri,
  A.~Nakata, J.~E. Pask, C.~Plessl, L.~E. Ratcliff, R.~M. Richard, M.~Rossi,
  R.~Schade, M.~Scheffler, O.~Sch{\"u}tt, P.~Suryanarayana, M.~Torrent,
  L.~Truflandier, T.~L. Windus, Q.~Xu, V.~W.-Z. Yu, D.~Perez, Roadmap on
  electronic structure codes in the exascale era, Modelling and Simulation in
  Materials Science and Engineering 31~(6) (2023) 063301.

\bibitem{bethkenhagen2023properties}
M.~Bethkenhagen, A.~Sharma, P.~Suryanarayana, J.~E. Pask, B.~Sadigh, S.~Hamel,
  {Properties of carbon up to 10 million kelvin from Kohn-Sham density
  functional theory molecular dynamics}, Physical Review E 107~(1) (2023)
  015306.

\bibitem{zhang2019equation}
S.~Zhang, A.~Lazicki, B.~Militzer, L.~H. Yang, K.~Caspersen, J.~A. Gaffney,
  M.~W. D\"ane, J.~E. Pask, W.~R. Johnson, A.~Sharma, P.~Suryanarayana, D.~D.
  Johnson, A.~V. Smirnov, P.~A. Sterne, D.~Erskine, R.~A. London, F.~Coppari,
  D.~Swift, J.~Nilsen, A.~J. Nelson, H.~D. Whitley, Equation of state of boron
  nitride combining computation, modeling, and experiment, Physical Review B
  99~(16) (2019) 165103.

\bibitem{wu2021development}
C.~J. Wu, P.~C. Myint, J.~E. Pask, C.~J. Prisbrey, A.~A. Correa,
  P.~Suryanarayana, J.~B. Varley, Development of a multiphase beryllium
  equation of state and physics-based variations, The Journal of Physical
  Chemistry A 125~(7) (2021) 1610--1636.

\bibitem{sharma2021real}
A.~Sharma, P.~Suryanarayana, {Real-space density functional theory adapted to
  cyclic and helical symmetry: Application to torsional deformation of carbon
  nanotubes}, Physical Review B 103~(3) (2021) 035101.

\bibitem{ghosh2019symmetry}
S.~Ghosh, A.~S. Banerjee, P.~Suryanarayana, {Symmetry-adapted real-space
  density functional theory for cylindrical geometries: Application to large
  group-IV nanotubes}, Physical Review B 100~(12) (2019) 125143.

\bibitem{kumar2020bending}
S.~Kumar, P.~Suryanarayana, Bending moduli for forty-four select atomic
  monolayers from first principles, Nanotechnology 31~(43) (2020) 43LT01.

\bibitem{kumar2022bending}
S.~Kumar, P.~Suryanarayana, On the bending of rectangular atomic monolayers
  along different directions: an ab initio study, Nanotechnology 34~(8) (2022)
  085701.

\bibitem{codony2021transversal}
D.~Codony, I.~Arias, P.~Suryanarayana, Transversal flexoelectric coefficient
  for nanostructures at finite deformations from first principles, Physical
  Review Materials 5~(3) (2021) L030801.

\bibitem{kumar2021flexoelectricity}
S.~Kumar, D.~Codony, I.~Arias, P.~Suryanarayana, Flexoelectricity in atomic
  monolayers from first principles, Nanoscale 13~(3) (2021) 1600--1607.

\bibitem{bhardwaj2022strain}
A.~Bhardwaj, P.~Suryanarayana, {Strain engineering of Janus transition metal
  dichalcogenide nanotubes: an ab initio study}, The European Physical Journal
  B 95~(3) (2022) 59.

\bibitem{bhardwaj2021torsionala}
A.~Bhardwaj, A.~Sharma, P.~Suryanarayana, Torsional strain engineering of
  transition metal dichalcogenide nanotubes: an ab initio study, Nanotechnology
  32~(47) (2021) 47LT01.

\bibitem{bhardwaj2021torsionalb}
A.~Bhardwaj, A.~Sharma, P.~Suryanarayana, Torsional moduli of transition metal
  dichalcogenide nanotubes from first principles, Nanotechnology 32~(28) (2021)
  28LT02.

\bibitem{bhardwaj2023ab}
A.~Bhardwaj, P.~Suryanarayana, {Ab initio study on the electromechanical
  response of Janus transition metal dihalide nanotubes}, The European Physical
  Journal B 96~(3) (2023) 36.

\bibitem{bhardwaj2022elastic}
A.~Bhardwaj, P.~Suryanarayana, {Elastic properties of Janus transition metal
  dichalcogenide nanotubes from first principles}, The European Physical
  Journal B 95~(1) (2022) 13.

\bibitem{ghosh2016higher}
S.~Ghosh, P.~Suryanarayana, Higher-order finite-difference formulation of
  periodic orbital-free density functional theory, Journal of Computational
  Physics 307 (2016) 634--652.

\bibitem{suryanarayana2014augmented}
P.~Suryanarayana, D.~Phanish, {Augmented Lagrangian} formulation of
  orbital-free density functional theory, Journal of Computational Physics 275
  (2014) 524--538.

\bibitem{kumar2023kohn}
S.~Kumar, X.~Jing, J.~E. Pask, A.~J. Medford, P.~Suryanarayana, {Kohn--Sham}
  accuracy from orbital-free density functional theory via {$\Delta$-}machine
  learning, The Journal of Chemical Physics 159~(24).

\bibitem{kumar2024fly}
S.~Kumar, X.~Jing, J.~E. Pask, P.~Suryanarayana, On-the-fly machine learned
  force fields for the study of warm dense matter: application to diffusion and
  viscosity of {CH}, arXiv preprint arXiv:2402.13450.

\bibitem{pratapa2016spectral}
P.~P. Pratapa, P.~Suryanarayana, J.~E. Pask, {Spectral Quadrature method for
  accurate O(N) electronic structure calculations of metals and insulators},
  Computer Physics Communications 200 (2016) 96--107.

\bibitem{suryanarayana2013spectral}
P.~Suryanarayana, On spectral quadrature for linear-scaling density functional
  theory, Chemical Physics Letters 584 (2013) 182--187.

\bibitem{banerjee2016cyclic}
A.~S. Banerjee, P.~Suryanarayana, {Cyclic density functional theory: A route to
  the first principles simulation of bending in nanostructures}, Journal of the
  Mechanics and Physics of Solids 96 (2016) 605--631.

\bibitem{sharma2023calculation}
A.~Sharma, P.~Suryanarayana, Calculation of phonons in real-space density
  functional theory, Physical Review E 108~(4) (2023) 045302.

\bibitem{thapa2023assessing}
B.~Thapa, X.~Jing, J.~E. Pask, P.~Suryanarayana, I.~I. Mazin, Assessing the
  source of error in the {Thomas--Fermi--von Weizs{\"a}cker} density
  functional, The Journal of Chemical Physics 158~(21).

\bibitem{kumar2022accurate}
S.~Kumar, B.~Sadigh, S.~Zhu, P.~Suryanarayana, S.~Hamel, B.~Gallagher,
  V.~Bulatov, J.~Klepeis, A.~Samanta, Accurate parameterization of the kinetic
  energy functional for calculations using exact-exchange, The Journal of
  Chemical Physics 156~(2) (2022) 024107.

\bibitem{lei2022universal}
X.~Lei, A.~J. Medford, A universal framework for featurization of atomistic
  systems, The Journal of Physical Chemistry Letters 13~(34) (2022) 7911--7919.

\bibitem{zhang2023version}
B.~Zhang, X.~Jing, S.~Kumar, P.~Suryanarayana, {Version 2.0.0-M-SPARC:
  Matlab-simulation package for ab-initio real-space calculations}, SoftwareX
  21 (2023) 101295.

\bibitem{xu2020m}
Q.~Xu, A.~Sharma, P.~Suryanarayana, {M-SPARC: Matlab-simulation package for
  ab-initio real-space calculations}, SoftwareX 11 (2020) 100423.

\bibitem{huang2023formation}
P.-W. Huang, N.~Tian, T.~Rajh, Y.-H. Liu, G.~Innocenti, C.~Sievers, A.~J.
  Medford, M.~C. Hatzell, Formation of carbon-induced nitrogen-centered
  radicals on titanium dioxide under illumination, JACS Au 3~(12) (2023)
  3283--3289.

\end{thebibliography}


\end{document}